 \RequirePackage{fix-cm}
 \documentclass[aps, prx, reprint, superscriptaddress]{revtex4-2}
 
\usepackage{amsmath} % assumes amsmath package installed
\usepackage{amssymb}  % assumes amsmath package installed

\usepackage{xcolor}         % colors
\usepackage{latexsym}
\usepackage{amsmath}
\usepackage{amssymb}
\usepackage{times}
\usepackage{graphicx}
\usepackage{epsfig}
\usepackage{array}
\usepackage{flafter}
\usepackage{listings}
\usepackage{algorithm,algpseudocode}
\usepackage{ulem}
\usepackage{verbatim}
\usepackage{hyperref}
\usepackage{qtree}
\usepackage{mathrsfs,amsmath} 

\usepackage[a4paper, total={6in, 8.5in}]{geometry}

\usepackage{newtxmath}

\graphicspath{ {Figures/} }

\usepackage{setspace}
\usepackage{dcolumn}  % Align table columns on decimal point
\usepackage{bm}       % bold math

\newcommand{\Rho}{\mathrm{P}}

\newcommand{\Ex}[1]{Example~\ref{ex:#1}}
\newcommand{\ex}[1]{example~\ref{ex:#1}}

\newtheorem{definition}{Definition}
\newtheorem{theorem}{Theorem}

\newtheorem{ExampleDef}{Example}%[section]

\newcommand{\Example}[3]{
  \begin{list}{}{
      \setlength{\leftmargin}{0em}} % Indent everything by this amount
    \item                           % Group everything in one item
    \small                          % Use a smaller font size
    \begin{ExampleDef} \rm          % select roman
      {\it \hspace{-1ex} #1}       % Use[1ex] to break line
      #2                                % The actual stuff
      \hfill {\large \boldmath $\Box$}  % The box
      \label{ex:#3}                      % Label the example
    \end{ExampleDef}
  \end{list}}

\begin{document}

\title{Exact computation of quantum wave functions for nonlinear potentials}
\author{Winfried Lohmiller}
\author{Jean-Jacques Slotine}
\affiliation{Massachusetts Institute of Technology}
%\date{\today}

%and curved spaces

%\thispagestyle{empty}
%\pagestyle{empty}

%%%%%%%%%%%%%%%%%%%%%%%%%%%%%%%%%%%%%%%%%%%%%%%%%%%%%%%%%%%%%%%%%%%%%%%%%%%%%%%%

\ \ 

\begin{abstract}

Recent work  shows that the Schr\"odinger equation can be solved exactly based only on classical least action. The computation is based on first computing the multi-valued action solutions of the Hamilton-Jacobi equation, computing the classical propagated density along all stationary action   multipaths, and finally constructing the exact kernel and wave function based on these classical multi-valued quantities alone. The method requires that the propagated classical square root density along each stationary action path is harmonic, as is the case for basic examples such as e.g., the double-slit, quantum tunneling, or the Einstein-Podolsky-Rosen experiment. For more complex cases, residual Bohm-like quantum potentials along each stationary action path can be eliminated through a time-rescaling. 

This paper proposes an extended version of the above construction which avoids the need for time-rescaling even in the most general cases. This is achieved by introducing in the Hamilton-Jacobi equation a real viscosity term, which classically simply corresponds to replacing a lumped parameter model with a continuum model including a volumetric stress tensor.  This keeps the earlier construction unchanged, except that the real stationary action multipaths are now real stationary viscous action multipaths defined with respect to an associated modified Lagrangian. As in the earlier version, the construction extends straightforwardly to the special and general relativistic contexts, thus describing in a common language the quantum, classical, and relativistic cases. 

This classically-based contiuuum version is especially useful to compute the quantum waves of systems with nonlinear potentials. Examples such as the quantum nonlinear pendulum and the relativistic propagator of the Higgs boson illustrate the discussion. 

Note that the global Bohm potential term in the lumped Hamilton-Jacobi equation associated with a Madelung transform does not have a classical equivalent, while the volumetric stress tensor term in our viscous Hamilton-Jacobi equation directly represents the real paths of a classical deformed particle field.

%Note that the Bohm quantum potential of a lumped Hamiltonian system vanishes for a Hamiltonian continuum since it misses the internal classical volumetric stress of this continuum.
%\wl{This volumetric stress tensor is actually needed to deform a particle field, which cannot be modelled with a lumped Hamilton-Jacobi. Not including this stress tensor actually results to the non-classical Bohm quantum potential for deformed particle fields, which is avoided by the proposed approach.}

\ \

\end{abstract}

\maketitle

%%%%%%%%%%%%%%%%%%%%%%%%%%%%%%%%%%%%%%%%%%%%%%%%%%%%%%%%%%%%%%%%%%%%%%%%%%%%%%%%
\section{Introduction}

Richard Feynman's path-integral formulation of quantum mechanics~\cite{Feynman48} is a central tool among results aimed at bridging classical and quantum physics. Recent work \cite{LohmillerSlotine2026QuantumWaves} shows that the Schr\"odinger equation can be solved exactly based only on classical viscous actions, in essence replacing all the zig-zag paths in Feynman's construction by a much sparser set of viscous action paths.  We call $\varphi$ the viscous action associated with the viscous Hamilton–Jacobi p.d.e of a continuum, by analogy with Hamilton’s action $\phi$ for the inviscid lumped Hamilton-Jacobi p.d.e. We show that a viscous action field $\varphi({\bf q},t )$ is a combination of the classical Lagrangian action field $\phi({\bf q},t )$ and the classical density field $\rho({\bf q},t )$.  

The original approach in \cite{LohmillerSlotine2026QuantumWaves, LohmillerSlotine2026BohmPotential} was strictly defined for a propagated action, density and kernel from an initial position or momentum to enable a harmonic propagated density. If this was not the case (as for hydrogen) a time-scaling was applied. We will see that the usage of a viscous action $\varphi$ removes this constraint and enables an efficient wave computation for non-linear Hamiltonians or curved inertia tensors. The question of efficient wave computation is thus shifted to that of efficient viscous action computation, which is the subject of this paper. 

Recall that the classical motion of a lumped physical system corresponds to a local extremum over variational paths ${\bf q} = (q^1, ..., q^N) \in \mathbb{R}^N$ of the system's Lagrangian action,  
%{\bf q}_o \veebar {\bf p}_o
\begin{eqnarray}
    \phi({\bf q}, t)  &=& \underset{{\bf q}(t)}{\rm{extremum}} 
   \int_o^t L(\dot{\bf q}(\theta), {\bf q}(\theta),\theta)  d \theta  \label{eq:LagrangianAction}  \\
   L &=& \frac{m}{2}  \dot{\bf q}^T {\bf M} \dot{\bf q} +  {\bf A}^T {\bf Q}  \dot{\bf q}  - V 
    \label{eq:Lagrangian} 
\end{eqnarray}
with final position ${\bf q}$ at time $t$, 
%initial position ${\bf q}_o$ or (exclusive $\veebar$) initial momentum ${\bf p}_o \ $,  
Lagrangian $L$, metric ${\bf M}({\bf q})$ which is multiplied with the mass $m$, potential energy $V({\bf q})$, vector potential ${\bf A}({\bf q})$, diagonal charge ${\bf Q}$ matrix with the constant charge $Q$ of each particle on the diagonal, see e.g. \cite{Feynman, Lagrange, Lovelock}. In this paper, unless otherwise specified, we will simply use the term action to refer to local stationary action. 

The symmetric metric ${\bf M}({\bf q})$ is required to be uniformly invertible, but is not necessarily positive 
definite. We use the standard covariant derivative and Laplace-Beltrami tensor operator ~\cite{Lovelock} of a covariant vector ${\bf f}({\bf q}, t) = (f_1, ..., f_N) \in \mathbb{R}^N$ or a scalar $f({\bf q}, t) $
\begin{eqnarray}
\nabla_{\bf M} \ {\bf f} &=& \frac{\partial {\bf f}}{\partial {\bf q} } - \sum_{h=1}^N \gamma_{jk}^h f_h \ \ \ \nabla_{\bf M} \cdot {\bf f} = Tr(\nabla_{\bf M} \ {\bf f}) \nonumber \\
 \Delta_{\bf M}  \ f &=& \nabla_{\bf M} \cdot \left(  {\bf M}^{-1} \frac{\partial f}{\partial {\bf q}}  \right) \label{eq:Delta} 
\end{eqnarray}
for a given metric ${\bf M}({\bf q})$ and Christoffel term \cite{Christoffel1869} $\sum_{m=1}^N \gamma_{jk}^m M_{lm} = \frac{1}{2} \left(\frac{\partial M_{jl}}{\partial q^k} + \frac{\partial M_{kl}}{\partial q^j} - \frac{\partial M_{jk}}{\partial q^l} \right)$.  The vector potential ${\bf A}({\bf q},t)$ is assumed to follow the Coulomb or Lorenz gauge $\nabla_{\bf M} \cdot {\bf A} = 0 \ $ \cite{feynmanquantum1998}.  

%The Lagrangian action $\phi({\bf q}, t)$ can be computed from $\frac{d \phi}{dt } =  L$ along a  path with with piecewise  ${\bf C}^2$ Lagrangian $ L \ $.

In this paper, we assume that $V \ge 0$ has a non-constant minimum $ V = 0 $ at an equilibrium point, and we separate variables using $\ \phi = E t + \Phi({\bf q})\ $, with $E$ a constant energy offset. The lumped Hamilton-Jacobi \cite{Goldstein, Hamilton, Jacobi} is given by
\begin{equation}
- {\frac {\partial \phi}{\partial t}} = H = {\frac {1}{2m}} \left( \frac{\partial \phi}{\partial {\bf q}} - {\bf Q}  {\bf A} \right)^T  {\bf M}^{-1}  \left( \frac{\partial \phi}{\partial {\bf q}} - {\bf Q}  {\bf A} \right) + V  \label{eq:H}
\end{equation}
A solution $\Phi$ always exists when the gauge-covariant momentum $  \frac{\partial \Phi}{\partial {\bf q}} - {\bf Q}  {\bf A}$  of (\ref{eq:H})  is imaginary~\cite{evans2010partial}. Note that a constant $V$ corresponds to a free particle in a possibly constrained space,  a case which was already studied  in \cite{LohmillerSlotine2026QuantumWaves} for the double slit, Aharonov-Bohm, potential well, tunneling, EPR, and QED examples,  and in the derivations of the Pauli, Dirac, and Maxwell equations.

If the Hamiltonian $H$ of (\ref{eq:H}) is not lumped but rather continuous and compressible, then the lumped Hamilton-Jacobi p.d.e. (\ref{eq:H}) turns in the viscous Hamilton-Jacobi p.d.e. \cite{evans2010partial}, also called viscous Bernoulli equation  \cite{Joseph2003ViscousPotentialFlow} \cite{prandtl1990fuehrer} \cite{Bernoulli1738} or Burger's~\cite{Burgers1948} p.d.e. (where the latter two use pressure = energy per volume element),
\begin{eqnarray}
- {\frac {\partial \varphi}{\partial t}} &=& H -  \nu i \Delta_{\bf M} \varphi \label{eq:viscHJ} \\
H &=& {\frac {1}{2m}} \left( \frac{\partial \varphi}{\partial {\bf q}} - {\bf Q}  {\bf A} \right)^T  {\bf M}^{-1}  \left( \frac{\partial \varphi}{\partial {\bf q}} - {\bf Q}  {\bf A} \right) + V  \nonumber
\end{eqnarray}
We separate again variables $\varphi = E t + \varPhi ({\bf q})$, where according to \cite{evans2010partial} a solution $\varPhi$ always exists where  gauge-covariant momentum $  \frac{\partial \varPhi}{\partial {\bf q}} - {\bf Q}  {\bf A}({\bf q})$  of (\ref{eq:viscHJ}) is imaginary.  As a result, a real kinematic viscosity $\nu$ implies a real contracted volumetric strain rate tensor $i \Delta_{\bf M} \varphi$, leading to a real volumetric stress $\nu i \Delta_{\bf M} \varphi$ in the continuum \cite{prandtl1990fuehrer}. Note a pressure term does not exist here, similarly to the theory of thin gases where atoms are modelled indvidually \cite{prandtl1990fuehrer}.

As a result, the related viscous Euler-Lagrange equation $\frac{d}{dt} \frac{\partial L}{\partial \dot{\bf q}} - \frac{\partial L}{\partial {\bf q}} = \nu i \Delta_{\bf M} \varphi$ stays real, i.e. we get a real path. 
(\ref{eq:viscHJ}). Hence by classical physics we mean any real, complex or quaternion solution of (\ref{eq:H}) or (\ref{eq:viscHJ}), which leads to a real path, noting that only the path and not the action can be measured. Quaternion momenta \cite{Goldstein} are actually standard in classical physics to describe rotations. 

The real volumetric stress $i \nu \Delta_{\bf M} \varphi \ne 0 $ is relevant in continuum mechanics in spatial regions which undergo rapid changes of volume, analogously to the case when a electron passes through a infinitesimal slit. Only if the volumetric stress is position independent, then the Bohm-quantum  potential of the lumped Hamilton-Jacobi (\ref{eq:H}) is zero \cite{LohmillerSlotine2026QuantumWaves, LohmillerSlotine2026BohmPotential}, which is e.g. the case for a free particle with a linear momentum. 

Finally, note that it is known in quantum physics (e.g., ~\cite{Chavanis2017}) that (\ref{eq:viscHJ}) is equivalent to the Schrödinger equation through an inverse Hopf-Cole transformation \cite{evans2010partial}. However this result is non-classical, since it is assumed that the stress is imaginary and the associated path is complex. By constrast,  we use in our classically-based construction an {\it imaginary} stationary action in the continuum, which implies that the velocity fields, the multipaths, and the volumetric stresses are all {\it real}. Furthermore, similarly to~\cite{LohmillerSlotine2026QuantumWaves}, our construction of the global wave includes all stationary action branches (which earlier interpretations~\cite{Chavanis2017} do not consider). 
%\jjs{  In this setting, (\ref{eq:viscHJ}) is a fully classical and real description of a deformable continuum. }
%does not contradict the classical lumped physics (\ref{eq:H}), although its rate-of-strain is imaginary and its position dynamics complex. 
%However, so far (5) was considered to be in contradiction to the classical lumped physics (4), 
%with its ’imaginary’ and hence non-classical rate-of- strain and ’complex’ position dynamics

In line with \cite{LohmillerSlotine2026QuantumWaves}, we will see that viscous action $\varphi$ can be decomposed into the Lagrangian action
\begin{equation}
\frac{d \phi}{dt } =  L  \label{eq:phi_dot=L}
\end{equation}
and compressible fluid density $\rho$
\begin{eqnarray}
    0 &=&  \frac{\partial}{\partial t} \ \rho + \nabla_{\bf M} \cdot (\rho \ {\dot{\bf q}} ) = \frac{d \rho }{dt}  + \rho \  \Delta_{{\bf M}}  \varphi 
    %= \frac{d \rho }{dt}  + \rho \  T \ \Delta'  \varphi \ \ \ \ \ \ 
    \label{eq:continuity} 
\end{eqnarray} 
from the classical Euler continuity o.d.e. \cite{Euler}, both computed along the viscous and not lumped dynamics $\dot{\bf q} = {\bf M} \nabla \varphi$. The associated multi-valued viscous action, action and density computation is summarized in section \ref{col} ~\cite{LohmillerSlotine2026QuantumWaves}.

As detailed in \cite{LohmillerSlotine2026QuantumWaves, LohmillerSlotine2026BohmPotential} and summarized in Section~\ref{quan}, Feynman's infinity of non-classical zig-zag paths of the Hamiltonian (\ref{eq:viscHJ}) can be reduced to the subset of all  classical viscous action paths
%, each weighted with the classical  propagated  square root density computed along each path. 
This yields an exact construction of the wave $\psi(t,{\bf q})$  of the Schr\"odinger equation 
\begin{equation}
\left[ \frac{\hbar}{i} \frac{\partial}{\partial t} + \frac{1}{2 m}\left( \frac{\hbar}{i} \nabla_{\bf M} - {\bf Q} {\bf A} \right) \cdot {\bf M}^{-1} \left( \frac{\hbar}{i} \nabla -  {\bf Q}  {\bf A} \right) + V  \right] \psi   \label{eq:Schroed} 
\end{equation}
  
Section \ref{nonlinearaction} of this paper systematically and exactly computes the viscous action for nonlinear potentials $V ({\bf q})$ where the volumetric strain rate $\Delta_{\bf M} \varphi ({\bf q},t)$ is indeed space dependent. This is a key improvement over approximate perturbation theory \cite{Heisenberg1925, Schrodinger1926, Wilson1929}, using the WKB approximation \cite{Wentzel1926, Kramers1926,Brillouin1926} of partial differential equations, or semi-classical approximations \cite{landauquantum, Maslov1972}. Note in general, $N>1$ dimensional systems first have to be decoupled  into $N$ separate dynamics using separation of variables or more generally the modular decomposition of nonlinear Hamiltonians recently derived in \cite{lohmiller2025contraction20naturalmetrics}. %Alternatively, $N>1$-dimensional holomorphic complex or quaternion systems can be transformed to harmonic action branches in scaled time.

Section \ref{examples} illustrates the approach with the harmonic and nonlinear pendulum oscillators and finally with a spatial decompostion of the two-slit experiment. Concluding remarks and perspectives are offered in Section~\ref{concluding}.

\section{Computing quantum waves exactly from viscous action} \label{secton2}

Rather than being based on a approximation or a mathematical expansion, our mapping in~\cite{LohmillerSlotine2026QuantumWaves} from classical to quantum mechanics relies on a different and exact mechanism. This section summarizes this construction, with the reader referred to \cite{LohmillerSlotine2026QuantumWaves}  for details.

\subsection{\label{col} Multi-valued local viscous actions and multipaths} 

Let us first introduce spatial inequality constraints on a multiply connected manifold, and then define viscous action branches on this manifold.
\begin{definition} 
{\it The constrained multiple connected manifold $ \ \mathbb{G}^N({\bf q} \ \veebar \ {\bf p}, t) \subseteq \mathbb{R}^N$ is defined by the $g=1, ..., G$ inequality constraints $ \ f_g({\bf q}, t)\le 0\ \  \veebar \ \ f_g({\bf p}, t)  \le  0 \ $.
The set of active constraints $\mathbb{G}  \subseteq  \{1, ..., G \}$  is the set of indices $g$  on the boundary $\partial \mathbb{G}^N$ of $\mathbb{G}^N$ such that $ \ f_g({\bf q}, t) =0 \ \ \veebar \ \ f_g({\bf p}, t) = 0 \ $.}
 \label{def:equalconstraint}
\end{definition}

\begin{definition}
{\it The viscous action branch set is the set of local viscous action fields $ \varphi_j({\bf q}, t), %\ j=1, ..., J, 
\ {\bf q} \in \mathbb{G}^N, t \ge 0$ which are different  at least at one $({\bf q}, t)  \ $, except for the integration constant for each given initial condition in (\ref{eq:LagrangianAction}).}
\label{def:branchset}
\end{definition}

Multi-valued viscous action branches naturally arise from given initial conditions in ${\bf q}_o$ or $\dot{\bf q}_o$ in (\ref{eq:LagrangianAction}) or at branch points. 
\begin{definition}
{\it The set of branch points $\mathbb{B}^N\subseteq  \mathbb{G}^N$ consists of all points where the set of distinct  $ \ \varphi_j({\bf q}, t)$ changes, for $t \ge 0 \ $. Each branch point ${\bf q}$ occurs $(b_{\bf q}-1)$ times in $\mathbb{B}^N$, where $ \ b_{\bf q} \in \mathbb{N} \ $ is the number of distinct viscous action solutions in the local neighborhood of ${\bf q}$.}
\label{def:branch}
\end{definition}

The Lagrangian collision forces $\lambda_g \frac{\partial f_g}{\partial {\bf q}}$ are activated by inequality constraints. The constrained Burger's equation can be derived similarly to an unconstrained Euler-Lagrange equation by local extremization \cite{Lagrange}  
\begin{eqnarray}
\delta \varphi  &=&  \int_o^t  \frac{\partial L}{\partial \dot{\bf q}} \ \delta \dot{\bf q} +  \frac{\partial (L - i \nu  \Delta_{\bf M} \varphi)}{\partial {\bf q}} \ \delta {\bf q} \ d \theta  \nonumber \\
&=& \left[   \frac{\partial L }{\partial {\bf q}}  \ \delta {\bf q}  \right]_o^t - \int_o^t \left( \frac{d}{dt} \frac{\partial L}{\partial \dot{\bf q}} - \frac{\partial L}{\partial {\bf q}} + i \nu \nabla \Delta_{\bf M} \varphi \right) \delta {\bf q} \ d \theta \nonumber \\
&=& \int_o^t \sum_{g \in \mathbb{G}}  \lambda_g \frac{\partial f_g}{\partial {\bf q}} \ \delta {\bf q} \ d\theta \nonumber 
\end{eqnarray}
where the Lagrange parameter $\lambda_g$ defines the magnitude of the cost gradient at the active constraint. The term in brackets is zero since $\delta {\bf q}$ is zero at the start and end points. In between $\delta {\bf q}$  can take on any arbitrary value.  A viscous action solution hence satisfies Burger's equation \cite{Burgers1948}
\begin{equation}
   \frac{d}{dt} \frac{\partial L}{\partial \dot{\bf q}} - \frac{\partial L}{\partial {\bf q}} = \frac{d \nabla \varphi_j}{dt}  +  \frac{\partial H}  {\partial {\bf q}}\  =  \sum_{g \in \mathbb{G}} \lambda_g \frac{\partial f_g}{\partial {\bf q}} \ - \ i \nu \ \nabla \Delta_{\bf M}  \varphi   \ \ \ \ \ \label{eq:EulerLangrange}
\end{equation} 
Note that, for $V \ge 0$ with a non-constant minimum $V = 0$,  the term  $- i \nu  \Delta_{\bf M} \nabla \varphi $ is real since $\varphi$ is imaginary. Also, in the optimization above, the volumetric strain rate $i \Delta_{\bf M}  \varphi$ acts as a real potential energy rather than as dissipation.

Summarizing \cite{LohmillerSlotine2026QuantumWaves} and the above, and introducing for generality an ensemble $\mathbb{E}$ of possible initial density conditions with given probabilities, leads to the following result. 
\begin{theorem}
The  multi-valued complex viscous action field $ \ \varphi_j({\bf q}, t) = E t + \varPhi ({\bf q})$ of the viscous Hamilton-Jacobi p.d.e (\ref{eq:viscHJ}) locally extremizes $L - i \nu  \Delta_{\bf M} \varphi$ (\ref{eq:Lagrangian}) on the constrained multiply connected manifold ${\bf q}, {\bf p} \in \mathbb{G}^N$ of Definition \ref{def:equalconstraint}, yielding the multipaths
\begin{eqnarray}
   \frac{d \nabla \varphi_j}{dt}  +  \frac{\partial H}  {\partial {\bf q}}\  &=&  \sum_{g \in \mathbb{G}} \frac{\partial f_g}{\partial {\bf q}} \lambda_g + \nu i \ \nabla \Delta_{\bf M} \varphi_j \ \ \ \ \ \label{eq:dotp} \\
   {\bf M}({\bf q})\ \frac{d{\bf q}}{dt}  &=& \nabla \varphi_j - {\bf Q} \ {\bf A} \label{eq:dotq} 
\end{eqnarray}
with the kinematic viscosity $\nu$ and where the constraint force $\lambda_g \ $ fulfills Definition \ref{def:equalconstraint}. 

The paths ${\bf q}$ are real for real initial or final conditions if $V$ is lower bounded, where we assume without loss of generality that $V = 0$ at this minimum.
\begin{itemize}
    \item For a non-constant $V$, the volumetric stress  $\nu i \nabla \Delta_{\bf M} \varphi$ is real. 
    \item For a constant $V$ in a flat space \cite{Lovelock}, the volumetric stress  $ \ \nu i \nabla \Delta_{\bf M} \varphi \ $ is zero, and thus the lumped Hamilton-Jacobi euqations apply. This was extensively studied in \cite{LohmillerSlotine2026QuantumWaves} for a position independent strain $\Delta_{\bf M} \varphi_j(t)$,   
\end{itemize}

The viscous action is said to be $\mathbb{B}$-valued, where $\mathbb{B}$ indexes the set $\mathbb{B}^N$ which accounts for all branch points $\mathbb{B}^N$ of Definition \ref{def:branch}. Branch points exist only at unbounded  $\Delta_{\bf M} \varphi_j({\bf q} \in \mathbb{G}^N, t)$ or unbounded $\nabla \varphi_j({\bf q} \in \partial \mathbb{G}^N, t)$.

Combining (\ref{eq:continuity}), (\ref{eq:dotq}) the propagated classical square root density can be computed along each extremal path ${\bf q}_j(t)$, yielding  
\begin{eqnarray}
    \sqrt{\rho}_j  &=&  e^{-\frac{1}{2} \int_{0}^{t} \Delta_{\bf M} \varphi_j(\theta, t)  \ d \theta} \label{eq:density}
\end{eqnarray}
where the gauge $ \ \nabla_{\bf M} \cdot {\bf A} = 0 \ $ is used. In addition the Lagrangian action $\phi_j$ is defined in (\ref{eq:LagrangianAction}) along the viscous path (\ref{eq:dotq}).
\label{th:Hamilton}
\end{theorem}

Note for a continuum, (\ref{eq:dotp}) and (\ref{eq:dotq}) correspond to a spatial fixed infinitesimal volume element of the continuum. $\nu i \nabla \Delta_{\bf M} \varphi$ defines the internal volumetric stress of the the continuum. When $\nu$ is zero, (\ref{eq:dotp}) is just the lumped Hamilton-Jacobi equation. 

Also note that branch points in Theorem \ref{th:Hamilton} occur at an initial given position or momentum, at singularities, at constraints  but not at the final point. At a branch point, the momentum (or for a momentum branch point, the  position) is undefined, which implies a stochastic event at that point. Classical physics is deterministic except at branch points, where it is stochastic, a result which allows the bridge to quantum physics in the next section. This represents an alternative to the Copenhagen interpretation \cite{bohr1928quantenpostulat, Heisenberg1925, Schrodinger1926, heisenberg1958physics}, where the statistical decision occurs at the measurement, i.e., at the endpoint of the classical path. This result was first derived in Lemma 3.3 in  \cite{LohmillerSlotine2026QuantumWaves}.

\subsection{Exact wave computation from classical propagated multi-valued action and density} \label{quan}

%As in \cite{LohmillerSlotine2026QuantumWaves} the wave computation is done in two steps introduced by Feynman~\cite{Feynman}, starting with a propagator computation and then integrating the propagator over all initial conditions. This is a key element in enabling the construction. It can be viewed more formally as writing the spectral decomposition of identity for the position or momentum operator~\cite{Schwartz1991analyse, Gelfand1964}.

The Schr\"odinger equation (\ref{eq:Schroed}) can be solved on each stationary branch $j \in \mathbb{B}$ of Definition (\ref{def:branchset}) 
%by the propagator or Feynman kernel \cite{Feynman} 
\begin{eqnarray}
    \psi({\bf q}, t) &=&  \sum_{j \in \mathbb{B}} \psi_j, \ \ \ \psi_j = e^{\frac{i}{\hbar} \varphi_{j}} \label{eq:viscpropagator} \\
    &=& \sum_{j \in \mathbb{B}} \sqrt{ \rho_j} \ e^{\frac{i}{\hbar} \phi_{j}} \label{eq:propagator} 
\end{eqnarray}
Equation (\ref{eq:viscpropagator}) computes the wave using the viscous actions $\varphi_{j}\ $ and kinematic viscosity $\nu = \frac{\hbar}{2m} \ $, with each of the summands  based on an inverse Hopf-Cole transformation \cite{evans2010partial} \cite{Broadbridge2015Burgers}. For each branch $j$, plugging the piecewise ${\bf C}^2$  $\psi_j({\bf x}, t)$ from  (\ref{eq:viscpropagator}) into the Schr\"odinger equation (\ref{eq:Schroed}) 
 {\scriptsize
 \begin{eqnarray} 
 \left[ \frac{\hbar}{i}  \ \frac{\partial }{\partial t} + \frac{1}{2m}\left( \frac{\hbar}{i} \nabla_{\bf M} -  {\bf Q} \ {\bf A}   \right) \cdot {\bf M}^{-1} \left( \frac{\hbar}{i} \nabla - {\bf Q} \ {\bf A} \right) + V \right] \ \psi_j &=& \nonumber \\
\left[ \frac{\partial \varphi_j}{\partial t}  + {\frac {1}{2m}} \ \left( \nabla \varphi_j- {\bf Q} \ {\bf A} \right)^T \ {\bf M}^{-1} \left( \nabla \varphi_j- {\bf Q} \ {\bf A} \right) + V - \frac{i \hbar}{2m} \Delta_{\bf M} \varphi_j \right] \ \psi_j  &=& 0 \nonumber 
\end{eqnarray}} 
exactly leads to the viscous Hamilton-Jacobi p.d.e. (\ref{eq:viscHJ}). 

Along each branch $j$, one has, from (\ref{eq:phi_dot=L}) and then (\ref{eq:density}),
\begin{eqnarray}
e^{\frac{i}{\hbar} \varphi_{j}}  \ = \ e^{\frac{i}{\hbar} \left[\int_{o}^{t}L \ - \  \frac{\hbar}{2i}\int_{o}^{t} \Delta \varphi_j(\theta)  \ d \theta \right] }  \ = \ \sqrt{ \rho_j} \ e^{\frac{i}{\hbar} \phi_{j}} \nonumber
\end{eqnarray}
Note that $\varphi_{j}$, $\phi_{j}$ and $\rho_{j}$ are computed along the paths (\ref{eq:dotq}) defined by the viscous case.

Let us summarize the above, including ~\cite{LohmillerSlotine2026QuantumWaves}, \cite{LohmillerSlotine2026BohmPotential}.
\begin{theorem} 
The wave
\begin{eqnarray}
  \psi({\bf q}, t) \ = \ \sum_{j \in \mathbb{B}} \psi_j, \ \ \ \psi_j = \sqrt{ \rho_j} \ e^{\frac{i}{\hbar} \phi_{j}}  \ = \   e^{\frac{i}{\hbar} \varphi_{j}} \ \ \ \ \ \ \ \ \ \ 
 \label{eq:Wave}
\end{eqnarray}
can be computed from the multi-valued viscous action field $\varphi_j = \varPhi_j ({\bf q}) - E_k t$ of the complex viscous Hamilton-Jacobi p.d.e. (\ref{eq:viscHJ}) of the ensemble $\mathbb{E} \ $ of Theorem \ref{th:Hamilton}. It can be expressed with the Lagrangian action $\phi_{j}({\bf q}, t)$ (\ref{eq:LagrangianAction}) and classical density fields (\ref{eq:density}) of Theorem \ref{th:Hamilton} which are computed along the multipaths (\ref{eq:dotq}) defined by the viscous case. 

The associated  quantum density matrix at time $t$
\begin{equation}
\varrho  ({\bf q}, t) \ =  \ \sum_{\epsilon \in \mathbb{E}} p^{\epsilon} \ \psi^{\epsilon} \psi^{\epsilon \dagger} \ \ \  \label{eq:Prob}
\end{equation}
is the determined forward mapping along all classical multipaths of Theorem \ref{th:Hamilton} from the initial quantum density distribution at $t=0$ for an ensemble $\mathbb{E} \ $,  with $\sum_{\epsilon \in \mathbb{E}} p^{\epsilon}  = 1$.
\label{th:quantum}
\end{theorem}
The above was complemented with Lemma 3.3 in  \cite{LohmillerSlotine2026QuantumWaves}, which classically derived the wave collapse at a measurement as an alternative to the Copenhagen wave collapse interpretation.
%Note that Theorems \ref{th:Hamilton} and \ref{th:quantum} immediately extend to complex or quaternion actions, as long as a real classical path can be constructed, for instance using superposition of the quadratic action in ${\bf q}'$ coordinates of Theorem \ref{th:NonlinearAction}. 

In contrast to  \cite{LohmillerSlotine2026QuantumWaves}, \cite{LohmillerSlotine2026BohmPotential},  the construction (\ref{eq:Wave}) above does not require a possible time-scaling for generality, as it relaxes the harmonic square root density assumption.
%The construction is always exact without time-scaling, even when the propagated densities along branches  of~\cite{LohmillerSlotine2026QuantumWaves}  involve non-zero Bohm-like potentials.
It also extends the propagated density / quantum propagator result in \cite{LohmillerSlotine2026QuantumWaves} to a general density and wave field.  
%In contrast to~\cite{LohmillerSlotine2026QuantumWaves}, it does not require Feynman kernels starting from Dirac density distributions of position. Thus, the sum in  (\ref{eq:Wave}) needs only be over the branch set $\mathbb{B}$, rather than on the product set $\mathbb{J}$ of initial positions with $\mathbb{B}$ as in~\cite{LohmillerSlotine2026QuantumWaves}.
Furthermore,
%The construction (\ref{eq:Wave}) of Theorem \ref{th:quantum}
\begin{itemize}
\item it is very different from the Madelung density \cite{Madelung1927} 
\begin{equation}
    \rho = |\psi|^2 \ \ \ \ \ \ \ \ \ \ \ \ \varphi = \hbar\,\arg(\psi) \label{eq:Madelung}
\end{equation}
which, in contrast to Theorem \ref{th:quantum}, is single valued and is based on a lumped
Hamiltonian (\ref{eq:H}) without volumetric stress of a continuum in (\ref{eq:dotp}). As a result it generates, in contrast to Theorem \ref{th:quantum}, for most waves a Bohm quantum potential \cite{Bohm1952}.
%, since the Madelung density does not account for the volumetric stress of a continuum in (\ref{eq:dotp}). 

%No Bohm-like quantum potential (which, in contrast to the viscosity term, has no analog in classical physics)  appears in either case.  Also note that the viscosity term describes the classical volumetric stress of a compressible continuum. 
\item It corresponds to the  Van Vleck \cite{VanVleck1928} kernel
\begin{equation} \ \psi \approx \sum_{\gamma}
\left|
\det\!\left(
-\frac{\partial^2 \varphi_\gamma}
{\partial \mathbf q\,\partial \mathbf q_o}
\right)
\right|^{1/2}
e^{ 
\frac{i}{\hbar}\varphi_\gamma
-\frac{i\pi}{2}\mu_\gamma } \label{eq:VanVlec}
\end{equation}
only if the propagated action $\varphi_\gamma( {\bf q},t, {\bf q}_o)$ is quadratic and single-valued, a $N \times (N -1) $ condition which implies that the pre-factor turns in a square root density. This is a much more restrictive condition than in Theorem \ref{th:quantum}, which is always exact for a Hamiltonian continuum or the original harmonic propagted square root density condition with possible time-scaling in \cite{LohmillerSlotine2026QuantumWaves}.

Also note that the Maslov phase index $\mu_\gamma$ is fundamentally distinct from the branch points of Theorem \ref{th:Hamilton}.  For instance, the harmonic oscillator has no branch points \cite{LohmillerSlotine2026QuantumWaves}, while the Maslov index counts the number of oscillations. Conversely, the periodic Maslov phase index cannot describe the 2 slits of the two-slit experiment, the $4$ main branches of the particle in the box or a $\uparrow$ or $\downarrow$ rotation for the hydrogen atom or spin in \cite{LohmillerSlotine2026QuantumWaves}. 

\item it only uses the $\mathbb{B}$-valued classical multipaths or actions from Theorem \ref{th:Hamilton}, which are a subset of all zig-zag paths in Feynman's path integral 
\begin{equation}
        \psi({\bf q}_o, {\bf q}, t)= \frac{1}{Z} \int_{{\bf q}_o}^{\bf q} e^{\frac{i  }{\hbar} \int_o^t L d \theta} \ \mathcal{D} {\bf q} \label{eq:Feynman}
\end{equation}
where $\mathcal{D} {\bf q}$ denotes the integration over $\infty^{\infty}$ stochastically time-sliced zig-zag paths and $Z$ is the normalization factor~\cite{Feynman}.

\item it exploits multi-valued complex or quaternion actions. In tunneling for instance, the relevant action cannot be real, but using a complex action with a real path solves the problem easily~\cite{LohmillerSlotine2026QuantumWaves}. Similarly, spin can be described using quaternion-based action~\cite{LohmillerSlotine2026QuantumWaves}, without requiring extensions of Feynman's zig-zag paths~\cite{Schulman1968PathIntegralSpin,Nielsen1988PathIntegralQuantizeSpin}. This was first introduced in \cite{LohmillerSlotine2026QuantumWaves}, e.g., for the lumped potential well, tunneling, Dirac, Pauli or Maxwell spin, using there an exact classical branch point definition in Theorem \ref{th:Hamilton} and the classical quantization Lemma 3.4.

\item the right side inverse Hopf-Cole transformation \cite{evans2010partial} is actually known in quantum physics (e.g., ~\cite{Chavanis2017}) for a single-valued action only. However this result is non-classical, since it assumes that the stress is imaginary and the associated path is complex. By constrast,  our classically-based construction uses an {\it imaginary} stationary action in the continuum, which implies that the velocity fields, the multipaths, and the volumetric stresses are all {\it real}. In addition, similarly to~\cite{LohmillerSlotine2026QuantumWaves}, our construction of the global wave includes all stationary action branches.
\end{itemize}

%The condition  $\Delta_{{\bf M}} \sqrt{\rho}({\bf q}, t) = 0$ of the propagated density is immediately verified in the following cases, with the first case (only) corresponding to a well-known result of Feynman~\cite{Feynman} and Van Vleck~\cite{VanVleck1928}.

The Bohm-quantum potential \cite{Bohm1952} is always zero in Theorem \ref{th:quantum}, in contrast to the Madelung density which does not account for the real volumetric stress $\nu i \nabla \Delta_{\bf M} \varphi_j$ (\ref{eq:dotp}) of a continuum. This was already demonstrated in \cite{LohmillerSlotine2026QuantumWaves} for non-quadratic action branches with position dependent $\Delta_{{\bf M}} \varphi_j({\bf q}, t)$, such as the conic action branches of the two-slit experiment, the nonlinear action branches of the Aharonov-Bohm experiment or the hydrogen atom. Section \ref{examples} will introduce further non-linear action examples where the Bohm-quantum potential exactly vanishes.

%in \cite{LohmillerSlotine2026QuantumWaves} , can still be handled exactly since the propagated square root density is harmonic on each branch. This is in contrast with the semi-classical results of \cite{VanVleck1928} or \cite{Feynman}, which require linear or quadratic actions to be exact since they miss the volumetric stress $\nu \nabla \Delta_{\bf M} \varphi$ in a continuum. 

As in~\cite{LohmillerSlotine2026QuantumWaves}, Theorem \ref{th:Hamilton} extends immediately to special and general relativity using Einstein's covariant relativistic Hamiltonian formulation \cite{Einstein,einstein1915field} as the Maxwell, Dirac or Klein-Gordon equations.

The following simple example illustrates the approach for a relativistic quadratic action in covariant form.
\Example{ Relativistic Propagator of Higgs Boson}{\ \ Consider a relativistic spin free particle with the relativistic covariant Hamilton-Jacobi p.d.e. or relativistic momentum energy relation (\ref{eq:viscHJ}),
\begin{equation}
0 \ = \ H \ = \ \frac{\partial \varphi}{\partial \bar{\bf q}}^T \frac{1}{m_o} {\bf M}^{-1} \frac{\partial \varphi}{\partial \bar{\bf q}}  - E_o, \ \ \   \bar{\bf q}  \in \mathbb{R}^4, \ \ \  \tau\ge 0,\ t \ge 0       \nonumber
\end{equation} 
with Cartesian relativistic position $\bar{\bf q} = (  t, {\bf q}^T)^T$, rest mass $m_o$, rest energy $E_o = m_o c^2$, proper time $\tau$ and Minkowski metric ${\bf M} =  {\rm diag}(c^2, -1, -1, -1)$.

The relativistic action 
\begin{equation}
\varphi  =  \frac{1}{2} \left( c^2 t^2 - ({\bf q} - {\bf q}_o)^T ({\bf q} -{\bf q}_o) \right) \nonumber    
\end{equation} 
solves (\ref{eq:viscHJ}) with $\Delta_{\bf M} \varphi = 4$, which implies that the propagated density (\ref{eq:density}) is independent of $\bar{\bf q}\ $. Hence Theorem \ref{th:quantum} yields the relativist wave (\ref{eq:Wave})
\begin{equation}
\psi =  e^{\frac{i }{\hbar} \varphi - 2 \tau} \nonumber
\end{equation}  
This wave represents an explicit integral of the relativistic Feynman propagator \cite{feynmanquantum1998, Peskin} and solves the Klein-Gordon equation \cite{Liboff}. This particle has no internal stress in Theorem \ref{th:Hamilton}.}{FreeRel}

%Finally the hydrogen atom in~\cite{LohmillerSlotine2026QuantumWaves} with position dependent $\Delta_{{\bf M}} \varphi_j({\bf q}, t)$ requires the density time scaling of Lemma \ref{lem:DeltTrafo}, leading to a harmonic propagated square root density. Also this case cannot be assessed with the quadratic action of \cite{VanVleck1928} or \cite{Feynman}. The goal of the next section is to systematically apply the time scaling of Lemma \ref{lem:DeltTrafo} for general nonlinear actions.

\section{Nonlinear action computation} \label{nonlinearaction}

To simplify the problem of finding an analytic action, this section assumes that a higher-dimensional viscous Hamilton-Jacobi dynamics is first decoupled in several $1$-dimensional viscous Hamilton-Jacobi equations, using either separation of variables or the recently developed modular decomposition of nonlinear Hamiltonians  \cite{lohmiller2025contraction20naturalmetrics}. 

Hence, we consider from now on a 1-dimensional, $j \in \mathbb{B}$-valued viscous Hamilton-Jacobi equation (\ref{eq:viscHJ}) of Theorem \ref{th:Hamilton}
\begin{eqnarray}
   - \frac{\partial \varphi_j}{\partial t} = H \ = \frac{\left( \nabla \varphi_j - Q A \right)^2 }{2 M}  +  V(q) - \frac{i \hbar}{2} \Delta_M \varphi_j \ \ \ \ \ \ \ \ \label{eq:visHJ1}
\end{eqnarray}
with metric $M(q)$, position $q\in \mathbb{G}^1$, vector potential $A(q,t)$, time $t \ge 0$, non-constant potential energy $\min (V) = 0$. 

The multi-valued viscous action of Theorem \ref{th:Hamilton} 
\begin{equation}
\varphi_j = \varPhi_j (q) - E_k t \ \ \ \ \ \ \ \ \ \  \varPhi_j = - i \phi_j - i \hbar \ \ln(\sqrt{\Rho}_j) \nonumber
\end{equation}
with the real constant eigenenergy $E_k$, dimensionless square root eigendensity $\sqrt{\Rho}_j(q)$, action $\phi_j = \int \sqrt{2 M V} dq + \int Q  A(q, t) dq\  $, solves (\ref{eq:visHJ1}) with  
\begin{eqnarray}
\nabla \varPhi_j &=& -i \nabla \phi_j - i \hbar \frac{\nabla \sqrt{\Rho}_j}{\sqrt{\Rho}_j} \nonumber \\ 
\frac{i \hbar}{2} \Delta_M \varPhi_j &=&  \frac{\hbar}{2} \Delta_M \phi_j - \frac{\hbar^2}{2M} \ \frac{(\nabla \sqrt{\Rho}_j)^2}{\sqrt{\Rho}_j^2} + \frac{\hbar^2}{2} \ \frac{\Delta_M \sqrt{\Rho}_j}{\sqrt{\Rho}_j} \nonumber \\
\frac{\left( \nabla \varPhi_j \right)^2 }{2 M} &=& - V -  \frac{\hbar}{M} \nabla \phi_j \   \frac{\nabla \sqrt{\Rho}_j}{\sqrt{\Rho}_j} -  \ \frac{\hbar^2}{2M} \frac{(\nabla \sqrt{\Rho}_j)^2}{\sqrt{\Rho}_j^2}
 \nonumber
\end{eqnarray}
and where the stationary classical square root eigendensity $\sqrt{\Rho}_j$ is a solution of the coordinate invariant Sturm-Liouville eigenvalue problem \cite{SturmLiouville1837} 
 \begin{equation}
 \hbar \Delta_M \sqrt{\Rho}_j + \ 2 \frac{\nabla \phi_j}{M} \nabla \sqrt{\Rho}_j  + (2 \frac{E_k}{\hbar} +\Delta_M \phi_j) \sqrt{\Rho}_j = 0 \label{eq:SturmLiouville}
\end{equation}
The two boundary conditions in (\ref{eq:SturmLiouville}) can be chosen to find a solution. The eigenvalue $E_k$ has to be defined such that the overall unnormalized probability $\int_{q \in \mathbb{G}^1} \psi_j(q) \psi_j^\dagger (q) dq, \  \psi_j = \sqrt{ \rho_j} \ e^{\frac{i}{\hbar} \phi_{j}}$ is bounded \cite{Liboff}.

Let us now summarize the above.
\begin{theorem}
The $j \in \mathbb{B}$-valued action fields of the 1-dimensional, viscous Hamilton-Jacobi p.d.e. (\ref{eq:visHJ1}) are
\begin{equation}
 \varphi_j = E_k t + \varPhi_j ({\bf q}), \ \ \ \varPhi_j = -i \phi_j - i \hbar \ \ln(\sqrt{\Rho}_j) \ \ \ \ \ \label{eq:AnalyticAction}
\end{equation}
with gauge-covariant action $\phi_j(q) = \int \sqrt{2 M V} dq + \int Q  A(q, t) dq\  $, piecewise ${C}^2 \ $ non-constant potential energy $\min (V) = 0$, the square root eigendensity $\sqrt{\Rho}_j(q)$ and the constant eigenvalue $E_k$ in (\ref{eq:SturmLiouville}).

The two boundary conditions can be chosen to find a solution. The eigenvalue $E_k$ has to be defined such that the overall unnormalized probability $\int_{q \in \mathbb{G}^1} \psi_j(q) \psi_j^\dagger (q) dq, \  \psi_j = \sqrt{ \rho_j} \ e^{\frac{i}{\hbar} \phi_{j}}$ is bounded \cite{Liboff}, which is fulfilled for a sub-exponential density $\sqrt{ \rho_j}\ $.

This last condition
\begin{itemize}
    \item  is naturally fulfilled in a bounded space $\mathbb{G}^1$ outside the neighborhood of branch points with unbounded $\Delta_M \phi_j$. This means that on each side of the branch point a different eigenwave and eigenvalue $E_k$ can exist. 
    \item in an unbounded space, implies that $E_k$ has to be chosen such that $\lim_{\rightarrow \pm \infty} \sqrt{\Rho}_j$ is sub-exponential.
\end{itemize}

%The square root eigen-density $\sqrt{\Rho}_j$ can be computed explicitly with the o.d.e (\ref{eq:SturmLiouville}) from the equilibrimm point to the next classical branch point of Theorem \ref{th:Hamilton}. In the neighborhood of the equilibrium point the above corresponds to a Hermitian square root eigendensity $\sqrt{\Rho}_{k_j}$ of a harmonic oscillator with constant eigenvalue $E_{k_j} = E_k, k_j \in \mathbb{N}$ or of a free particle.

%\jjs{  For constant $V=0$ and \wl{$E = p_0^2 / m\ $}, the action field is with (\ref{eq:density})
%\begin{equation}
% \varphi_j = q p_o - E t, \ \ \ \sqrt{\rho}_j = e^{-\frac{1}{2} \int_{0}^{t} \Delta \varphi_j(\theta)(t)  \ d \theta}  \label{eq:AnalyticActionReal}
%\end{equation}
%with real constant momentum $p_0 = \sqrt{2 M V} $, square root density $\sqrt{\rho}_j\ $, and constant eigenvalue \wl{$E_k$}. The strain rate $\Delta \varphi_j(\theta)(t)$ may stem from a cross-coupling of the covariant derivative (\ref{eq:Delta}) for decoupled higher-dimensional systems. }
\label{th:NonlinearAction}
\end{theorem}
That the related computation of the last condition is straightforward, also in the nonlinear case, will be shown for the nonlinear pendulum in section (\ref{examples}).

Theorem \ref{th:Hamilton} can then be used to compute the multi-valued Lagrangian action $\phi_j$ and density $\rho_j$ and Theorem \ref{th:quantum} can be used to compute the related wave $\psi$.

Theorem \ref{th:NonlinearAction} extends the local perturbation theory \cite{Heisenberg1925, Schrodinger1926, Wilson1929}, using WKB or Taylor approximations \cite{Wentzel1926, Kramers1926, Brillouin1926} to a globally exact computation until and beyond the classical branch points of Thereom \ref{th:Hamilton}. Note that a Taylor series expansion of the Schr\"odinger p.d.e. only guarantees local convergence around the equilibrium point, since the next singularity is unknown. By contrast, the eigendensity o.d.e. (\ref{eq:SturmLiouville}) is exactly integrable to the next known classical branch point of Theorem \ref{th:Hamilton} with unbounded volumetric stress $\frac{\hbar}{2} \Delta_M \phi_j$. Naturally beyond this branch point another integration of the eigendensity o.d.e. \ref{eq:SturmLiouville} has to be performed.

Finally note that theorem \ref{th:NonlinearAction} is exact compared to the semi-classical approaches \cite{landauquantum, VanVleck1928, Maslov1972, BerryMount1972Semiclassical, Bohm1952}, {\it which all miss the real volumetric stress term $i \frac{\hbar}{2} \Delta_M \phi_j$ of a classical continuum} beyond the missing action branches.

\section{Examples} \label{examples}

The approach is first illustrated on the harmonic oscillator, whose solution is known,  and then applied to the nonlinear pendulum oscillator, where only approximate solutions are known so far. Finally, it is applied to the two-slit example to illustrate the multi-valued action and spatial separation technique for higher-dimensional systems.

All examples illustrate that the technique is straightforward, standardized and exact for nonlinear potential energies. While the actions in all examples, beyond the harmonic oscillator, are not quadratic, the computed waves are still exact. This is a key difference to the semi-classical Van Vleck \cite{VanVleck1928} or Feynman \cite{Feynman} method which are only exact for a single-valued real quadratic action since they miss, similarly to the Bohm potential or Madelung term the real volumetric stress of the continuum in Theorem \ref{th:Hamilton}.

\Example{Harmonic oscillator}{\ \ Consider the viscous Hamilton-Jacobi p.d.e. (\ref{eq:viscHJ}) of Theorem \ref{th:Hamilton}
\begin{eqnarray}
   - \frac{\partial \varphi}{\partial t} = H \ = \frac{1}{2} \left( \frac{\nabla \varphi^2}{m}   +  m \omega^2  q^2 \right) - \frac{i \hbar}{2 m} \Delta \varphi  \nonumber
\end{eqnarray}
with constant angular frequency $\omega$, mass $m$, real volumetric stress $\frac{i \hbar}{2} \Delta_m \varphi $ of the particle, Cartesian position $q \in \mathbb{R}$ and time $t \ge 0$.
The single-valued real viscous action of Theorem \ref{th:NonlinearAction} is 
\begin{equation}
\varphi = \varPhi (q) - E t, \ \ \ \varPhi = i \frac{m \omega}{2} q^2 - i \hbar \ \ln(H_k) \nonumber
\end{equation}
where the eigen square root eigen-density $\sqrt{\Rho}_k(q) = H_k, k \in \mathbb{N}$ is a Hermite polynomial \cite{Hermite1864}, which verifies (\ref{eq:SturmLiouville})
\begin{equation}
\Delta H_k - 2 m \omega q \nabla H_k + (2 E_k -1) H_k = 0 \nonumber
\end{equation}
The eigenwave is $\psi_k = H_k(q) e^{- q^2 / 2}$ for $m \omega / \hbar = 1$ according to Theorem \ref{th:quantum}. The Hermite polynoms are sub-expontial in $\lim_{\rightarrow \pm \infty} \sqrt{\Rho}_j$ for $E_k = k + \frac{1}{2}$.

Although the viscous stationary action $\varPhi $ is imaginary, the classical path is real,
\begin{equation}
    \ddot{q} = \Delta_m \varPhi \ \nabla \varPhi \nonumber
\end{equation}}{Harmonic}
We now turn to the simplest potential energies for which no exact analytic solution of the Schr\"odinger equation is known so far.

\Example{Nonlinear Pendulum}{ \ \ Consider the viscous Hamilton-Jacobi p.d.e. (\ref{eq:viscHJ}) of Theorem \ref{th:Hamilton}
\begin{eqnarray}
   - \frac{\partial \varphi}{\partial t} = H \ = \frac{1}{2} \left( \frac{\nabla \varphi^2 }{m}  +  m  \omega^2 (\sin q)^2 \right) - \frac{i \hbar}{2m} \Delta \varphi  \nonumber
\end{eqnarray}
with constant angular frequency $\omega$, constant mass $m$, real volumetric stress $\frac{i \hbar}{2m} \Delta \varphi $, periodic dimensionless Cartesian position $q \in [0; 2 \pi [$ and time $t \ge 0$. 
The nonlinear pendulum oscillator describes the phase of the Josephson junction in transmons~\cite{Martinis1987QuantumPhaseJosephson}.

The single-valued real viscous action of Theorem \ref{th:NonlinearAction} is 
\begin{equation}
\varphi = \varPhi ({\bf q}) - E t, \ \ \ \varPhi = i \frac{m \omega}{2} (\cos q-1) - i \hbar \ \ln(\sqrt{\Rho}_k) \nonumber
\end{equation}
where the square root eigendensity $\sqrt{\Rho}_k$ verifies (\ref{eq:SturmLiouville})
\begin{equation}
\frac{\hbar}{m \omega} \Delta \sqrt{\Rho}_k - 2  \sin q \nabla \sqrt{\Rho}_k + (\frac{E_k}{\omega \hbar} - \cos q  ) \sqrt{\Rho}_k = 0 \label{eq:eigwavepend}
\end{equation}
%With the eigenenergies $E_k = \hbar \omega (k+\frac{1}{2}), \  k \in \mathbb{N}$ of the local harmonic oscillator around the equilibrium points $q = 2 \pi l, l k \in \mathbb{Z}$ we can rewrite the above as 
which is normalized with $\frac{\hbar}{m \omega} = 1$ as

We impose the constraint that the eigendensity is $2 \pi$ periodic, which implies an identical eigendensity and slope at $q = 2k \pi, k \in \mathbb{Z}$. The normalized $\frac{E_k}{\omega \hbar}$ then has to be chosen to assure this periodicity, e.g., using the shooting method. The resulting square root eigendensity of (\ref{eq:eigwavepend}) is shown in Figure \ref{fig:pendulumeigendensity}. It is sub-exponential in $q \in [0; 2 \pi [$ as required by Theorem \ref{th:quantum}.
\begin{figure}
\centering
    \includegraphics[scale=0.45]{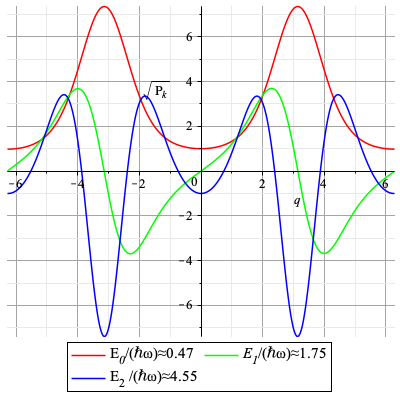}
    \caption{Periodic pendulum square root eigendensities} 
    \label{fig:pendulumeigendensity}
\end{figure}

The $2 \pi$ periodic eigenwave is according to Theorem \ref{th:quantum} and \ref{th:NonlinearAction}
\begin{equation}
  \Psi_k({q}) \ = \ \sqrt{ \Rho_k} \ e^{\frac{1}{2} (\cos q -1)}  \nonumber
  %\ = \  \sum_{j \in \mathbb{B}}  e^{\frac{i}{\hbar} \varphi_{j}}  
\end{equation}
whose un-normalized eigenwaves  are shown in Figure \ref{fig:pendulumeigendensity}.
\begin{figure}
\centering
    \includegraphics[scale=0.45]{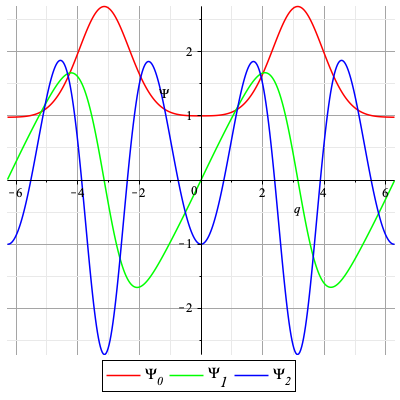}
    \caption{Periodic pendulum eigenwaves} 
    \label{fig:pendulumeigenwave}
\end{figure}
%One can see that locally around $q=2 \pi l$ it corresponds to the Hermitian polynomials  which around $|q| + 2 \pi l >1$ becomes non polynomial.

Although the viscous stationary action $\varPhi $ is imaginary, the classical path is again real
\begin{equation}
    \ddot{q} = \Delta_m \varPhi \ \nabla \varPhi \nonumber
\end{equation}}{Sinus}

The last example illustrates that solving the viscous Hamilton-Jacobi is straighforward for nonlinear Hamiltonians in contrast to solving the  Schr\"odinger equation.

The following two-slit example illustrates the importance of the multi-valued actions and how spatial decomposition can be applied to reduce a $2$-dimensional Hamiltonian in $2$ one-dimensional Hamiltonians. This result with $V=0$, first derived in \cite{LohmillerSlotine2026QuantumWaves}, is now based on the standardized technique developed here.

\Example{Two-slit experiment}{\ \ Consider the Hamilton-Jacobi p.d.e. (\ref{eq:viscHJ}) of Theorem \ref{th:Hamilton} on the two-connected manifold $\mathbb{G}^3$ 
\begin{eqnarray}
- {\frac {\partial \varphi_j}{\partial t}} &=& H  =  \frac{1}{2 m} \nabla \varphi_j^T \nabla \varphi_j \nonumber \\ 
{\bf q} &=& (q^1, q^2, q^3)^T \in \mathbb{G}^3 = \mathbb{R}^3 \setminus \mathbb{W}^3 \nonumber  
\end{eqnarray}
with normlized position ${\bf q}$, constant mass $m$, wall  $ \ \mathbb{W}^3 = \{ q^1 = 0  \}  \setminus  \mathbb{H}^3$ with two holes  $\mathbb{H}^3 = \{ {\bf q}_1, {\bf q}_2 \}, {\bf q}_{1, 2} = (0, \pm 5, 0)^T $, and initial momentum $p_0$ of the particle (for instance, an electron). 

1. In front of the wall ($q^1 < 0$) one has a free particle with action ${\varphi}_j = p_0 \ q^1$  and constant density of Theorem \ref{th:Hamilton}.

2. Behind the wall $q^1  \ge 0$, the $\mathbb{B} = \{1, 2 \}$ - valued actions of Theorem \ref{th:NonlinearAction} are with $E = \frac{p_0^2}{2 m}$, the radius $r_j = \sqrt{ ({\bf q} - {\bf q}_j)^T({\bf q} - {\bf q}_j)}$ and the central angle $\theta_j$  
\begin{equation}
{\varphi}_j  = E t + p_0 \ r_j  \nonumber
%+ \varPhi_j, \ \ \ \varPhi_j =  i \hbar \ \ln(r_j / r_o)  
\end{equation}

The dynamics is in radius $r_j$ and solid angle $\theta_j$ fully decoupled with the gradient and covariant derivative (\ref{eq:Delta}) in the spherical metric  
\begin{equation}\nonumber
\nabla \varphi_j = 
\begin{pmatrix}
p_0 \\
0 
\end{pmatrix} \ \ \ \ \ \ 
\nabla_{\bf M} \nabla \varphi_j = 
\begin{pmatrix}
0 & 0 \\
0 & p_0  r
\end{pmatrix} \ \ \ \ \ \ {\bf M} = 
\begin{pmatrix}
1 & 0 \\
0 & r^2
\end{pmatrix}
%\label{eq:densitytwoslit}
\end{equation}
The dynamics in $r_j$ has no strain rate. The dynamics in $\theta_j$ has no action and is only driven by a time-varying strain rate $p_0 / r_j(t)$, thanks to the decoupling. This is the reason why no position dependent stress has to be modeled in each decoupled  Hamiltonian dynamics. Both related classical paths are real with $\dot{\theta}_j = 0$, $\dot{r}_j = p_0/m$, and they fulfill (\ref{eq:viscHJ}).

Accordingly, the Lagrangian action and classical density of Theorem \ref{th:Hamilton} are
\begin{equation}
{\phi}_j  =
\left\{ 
\begin{array}{ll}
p_0 \ q^1 - E t  \\
p_0 \ r_j - E t   
\end{array}
\right. \ \ 
   \rho_j = 
\left\{ 
\begin{array}{ll}
1 & \text{  for  } q^1 < 0 \\
\frac{1}{r_j^2} %= \rho_o \frac{r_o}{r_o + \frac{p_0}{m}t}  
& \text{ for  } q^1 \ge 0  
\end{array}
\right.  
\end{equation}
Both slits are branch points with infinite classical volumetric stress and density in central angle direction $\theta_j$ and fully elastic collision forces in Theorem \ref{th:Hamilton} in this direction. 

The classical non-Lipschitz constraint forces (\ref{eq:dotp}) in the two slits lead to an infinity of radial paths (\ref{eq:dotq}) from each slit, which connect every measurement pixel on the screen at  with two least action paths (\ref{eq:dotq}). Thus, the statistical density distribution $\rho({\bf q})$ behind the slits is just the evolution of the Dirac densities in the slits along the density path integrals (\ref{eq:density}) after the slits. 
%The Taylor series expansion of Theorem \ref{th:NonlinearAction} is only valid on both sides of the branch points, i.e. the slits. %The Dirac approximation $  \ \hbar \ \Delta_{\bf m} \varphi \approx 0 \ $ of \cite{Dirac33} is not permissible since $\Delta_{\bf m} \varphi \ $ is actually unbounded in the slits.

Theorem \ref{th:quantum} in turn yields the (un-normalized) wave function (\ref{eq:Wave}), 
\begin{eqnarray}
{\psi} &=&  \sum_{j \in \mathbb{B}}  \sqrt{\rho_j} \ \ e^{\frac{i }{\hbar} \varphi_j} \nonumber \\
&=&  e^{-\frac{i}{\hbar} E t}
\left\{ 
\begin{array}{ll}
 e^{\frac{i}{\hbar} p_0 q^1 } & \ \text{for \ } q^1 < 0 \\
  \frac{1}{r_1}\  e^{\frac{i}{\hbar} p_0 r_1 } + \frac{1}{r_2}\ e^{\frac{i}{\hbar} p_0 r_2 }  & \ \text{for  \ } q^1 \ge 0  
\end{array}
\right.   \nonumber
\end{eqnarray}  
One can directly confirm that each term fulfills the Schr\"odinger equation (\ref{eq:Schroed}). 

The Hamilton-Jacobi formulation of Theorem \ref{th:Hamilton} cannot deterministically predict from the slits where the particle hits the screen. Indeed, in the slits only the position is given, but not the momentum direction, which is affected by the non-Lipschitz constraint force. However, the path from the slit to the screen is a determined straight line following (\ref{eq:dotq}) of Theorem \ref{th:Hamilton}. Theorem \ref{th:Hamilton}, in line with Lemma 3.3 in  \cite{LohmillerSlotine2026QuantumWaves}, thus allows a classically-based interpretation that the decision where the particle hits the screen is already taken in the slits from the non-Lipschitz constraint force (\ref{eq:dotq}) of Theorem \ref{th:Hamilton}, i.e.,  before the final position is measured on the screen.}{Twoslit}

Finally, note that neither the semi-classical Madelung (\ref{eq:Madelung}) equation, without Bohm-quantum potential,
\begin{eqnarray}
      \rho &=& |\psi|^2, \ \  \varphi = \hbar\,\arg(\psi) \nonumber \\
      \psi &=& \frac{1}{r_1}\  e^{\frac{i}{\hbar} p_0 r_1 } + \frac{1}{r_2}\ e^{\frac{i}{\hbar} p_0 r_2 } \nonumber
\end{eqnarray}
nor the Van Vleck equation (\ref{eq:VanVlec}) 
\begin{equation} \ \psi \approx \sum_{\gamma}
\left|
\det\!\left(
-\frac{\partial^2 \varphi_\gamma}
{\partial \mathbf q\,\partial \mathbf q_o}
\right)
\right|^{1/2}
e^{ 
\frac{i}{\hbar}\varphi_\gamma
-\frac{i\pi}{2}\mu_\gamma } \nonumber
\end{equation}
can compute this result exactly behind the wall, since the Van Vleck  pre-factor
$$
\frac{\partial^2 \varphi_\gamma}
{\partial \mathbf q\,\partial \mathbf q_o} \frac{1}{p_0} 
= \frac{\bf I}{r_j} -
\frac{({\bf q} - {\bf q}_j)({\bf q} - {\bf q}_j)^T}{r_j^{3}}.
$$
only approximates the classical density and does not fulfill the Schrödinger equation. In addition two slits cannot be described with a Maslov phase index. The density approximation of the Van Vleck pre-factor and the Maslov phase index approximation of classical branch points in Theorem \ref{th:Hamilton} (left/right turn in hydrogen or spin, 4 basic reflections in particle in the box, local branch points, ...) are the main reason why Van Vleck or Feynman cannot compute all the examples, or the Dirac, Pauli and Maxwell equations in \cite{LohmillerSlotine2026QuantumWaves} exactly, except for the harmonic oscillator. In addition, all these semi-classical techniques are limited to real-valued action, whereas Theorem \ref{th:Hamilton} allows complex actions, with real paths, to cover tunneling, spin or Hamiltonian continua with a lower bounded potential energy. 

Finally the wave of the nonlinear pendulum \ex{Sinus} was so far not computable exactly. This underlines the practicable relevance of this technique and illustrates why our approach is not semi-classical.

\section{Summary} \label{concluding}

The first point of this paper is to extend the exact quantum wave computation of lumped Hamiltonian systems in \cite{LohmillerSlotine2026QuantumWaves} to classical Hamiltonian continua. The multi-value action approach of \cite{LohmillerSlotine2026QuantumWaves} is a prerequisite. Existing techniques as Van Vleck's \cite{VanVleck1928} or Madelung's \cite{Madelung1927}, as well as \cite{LohmillerSlotine2026QuantumWaves}, use a lumped Hamiltonian model, in contrast to the continuum Hamiltonian model in Theorem \ref{th:Hamilton}. This is the main reason why \cite{LohmillerSlotine2026QuantumWaves} has to apply time-scaling for the general nonlinear case and why the other techniques are in general only approximate. The result in Theorem  \ref{th:quantum} is a direct extension of~\cite{LohmillerSlotine2026QuantumWaves}:
\begin{itemize}
    \item If the continuum has a strain rate, then the viscous Hamilton-Jacobi of Theorem \ref{th:Hamilton} has to be applied. Only if the strain rate $i \nabla \Delta_{\bf M} \varphi_j$ (\ref{eq:dotp}) is purely time-dependent does the Hamiltonian behave as lumped Hamiltonian. If the strain rate $i \nabla \Delta_{\bf M} \varphi_j$ (\ref{eq:dotp}) is position dependent, no Bohm-quantum potential occurs and no time rescaling in \cite{LohmillerSlotine2026QuantumWaves} has to be applied.

    \item The use of the inverse Hopf-Cole transformation (\ref{eq:Wave}) is well-known in quantum physics for a single-valued action. However, this result does not have a classical interpretation, since in the associated viscous Hamilton-Jacobi p.d.e. (\ref{eq:dotp}) the strain rate is imaginary and the associated path is complex. Using instead an imaginary action shows that (\ref{eq:dotp}) describes a classical continuum, extending the classical lumped physical description (\ref{eq:H}). %These non-classical \wl{ interpretations } are resolved in Theorem \ref{th:Hamilton} for lower-bounded potential energies, since here the action is imaginary.

\item The decomposition in classical density with Lagrangian action of \cite{LohmillerSlotine2026QuantumWaves} is reconfirmed in Theorem \ref{th:quantum} in the most general case. The multi-valued action of \cite{LohmillerSlotine2026QuantumWaves} and the Hamilonian continua modelisation are a pre-requisite for this formulation.\\

\item The Bohm-Madelung equation (\ref{eq:Madelung}) and the Van Vleck equation (\ref{eq:VanVlec}) do not describe the Hamiltonian as classical continua with a volumetric stress. This is one main reason beyond the non-consideration of branch points in Theorem \ref{th:Hamilton} and the non-usage of complex actions with real paths why they do not exactly correspond to a Hamiltonian system \cite{LohmillerSlotine2026QuantumWaves}.

\end{itemize}

\ \

\ \

The extension to a continuum Hamiltonian in  Theorem \ref{th:Hamilton} enables the second point of this paper, i.e., the introduction of Theorem \ref{th:NonlinearAction}, which computes for 1-dimensional, nonlinear systems the eigen-action and eigen-density exactly using the solution of an associated  Sturm-Liouville eigenvalue problem.  For higher-dimensional systems, the same approach can be used be first decoupling it into $N$ decoupled Hamiltonian dynamics.

The well-known harmonic oscillator is first used in \Ex{Harmonic} to illustrate the theory. The wave solution of the nonlinear pendulum in \Ex{Sinus} is new, and shows that conceptually the approach remains straightforward for general nonlinear potentials. In principle, this avoids the need for approximate perturbation theory \cite{Heisenberg1925, Schrodinger1926, Wilson1929}, the WKB approximation \cite{Wentzel1926} of partial differential equations, or semi-classical approximations \cite{landauquantum}. The two-slit experiment is used to show how the decoupling works for $2$-dimensional systems.

Current research focuses on applying Theorem \ref{th:quantum} and \ref{th:NonlinearAction} to the quartic oscillator, the Schwarzschild metric and the hydrogen fine structure, to further illustrate the use of the technique to compute exact quantum solutions based on classical nonlinear Hamiltonian continua.

%%%%%%%%%%%%%%%%%%%%%%%%%%%%%%%%%%%%%%%%%%%%%%%%%%%%%%%%%%%%%%%%%%%%%%%%%%%%%%%%
%\section{ACKNOWLEDGMENTS}

%This paper benefited from discussions with Alain Aspect, Michel Devoret, Steven Girvin, Serge Haroche, Christian Pehle, Carlo Rovelli, Stephan Schiller, and Pierre Rouchon. 

%%%%%%%%%%%%%%%%%%%%%%%%%%%%%%%%%%%%%%%%%%%%%%%%%%%%%%%%%%%%%%%%%%%%%%%%%%%%%%%%

\normalem
\bibliographystyle{abbrv}
\bibliography{References}{}

%\bibliographystyle{abbrv}
%\fontsize{10}{0} \selectfont \bibliography{References}{}}

\end{document}